**Andrea Scharnhorst**
**Data Archiving & Networked Services (DANS)**

**Richard P. Smiraglia**
**Institute for Knowledge Organization and Structure, Inc.**

# Chapter 1
# The Need for Knowledge Organization
## Introduction to the book *Linking Knowledge: Linked Open Data for Knowledge Organization*[†]

**Abstract**
This book is not restricted to semantic web (SW) technologies. An aspiration was to contribute to the awakening of a dialogue between information and documentation concerned with knowledge organization systems (KOSs), and branches in computer science with an emphasis on machines, algorithms and ontologies. The technological evolution of the last decades has not only fostered the emergence of ever more KOSs but also semantic web technologies. Both the actions of "making a KOS" and "applying existing KOSs" represent research. The design of an information layer for a knowledge domain and the design of a domain specific research process are intrinsically interwoven. We extended our intervention to KOS practices into education, by presenting a translation of existing standards and recommendations about linked open data (LOD) publishing for non-experts. The chapters describe the state of the art in providing KOSs as semantic artefacts; how the state of the art is applied in new fields; how the state of the art is pushed towards new technological solutions by being confronted with new applications; how best practices need to be tailored towards specific solutions; and what challenges occur when merging new and old ways of expressing KOSs. The linked data (LD) ecosystem represents a source of knowledge generation, acquisition, production and dissemination. The underlying discourse shows historical vision alongside the promise of linking knowledge for interaction. The already maturing ecosystems of the SW are interlocking information institutions clearly devoted to the expansion of human experience through the growth of knowledge interaction.

## 0.0 From the very beginning …
The title of this book is not arbitrary. While the monograph was produced in the context of a project about linked data (the Digging into the Knowledge Graph (DIKG) project[1]) the content of this book is not restricted to semantic web (SW) technologies. Instead, through the chapters, problems are addressed that prove to be almost eternal when it comes to the organization of knowledge.

From the very beginning the project—as now documented in this book's chapters— aspired to contribute to the awakening of a seemingly forlorn dialogue between those branches in the sciences of information and documentation that used to reflect about classifications, or knowledge organization systems (KOSs), and those branches in computer science that equally address the use of KOSs, but with a strong emphasis on machines, algorithms and ontologies.

[†] This work was funded by the European Commission T-AP Grant Agreement ID: 613167. We would like to acknowledge the collaboration with the VU Amsterdam Knowledge representation and reasoning group and Triply, an Amsterdam Startup. Part of this work started during visits of Richard Smiraglia at the Virtual Knowledge Studio (VKS-KNAW) and with the eHumanities group (KNAW), continued by his KNAW Visiting professorship grant and his fellowship with Data Archiving and Networked Services (DANS-KNAW).

In this introduction, we detail our motivation for why this discourse needs to be awakened and how best to do this. In doing this we rely on insights from the fields of knowledge organization (KO) and science and technology studies (STS). Most of our use cases stem from the digital humanities (DH). We also emphasise the role of visualization as means to support translation across or between different knowledge domains as part of the essential knowledge exchange.

## 1.0 Organizing knowledge, linking knowledge

Ordering of knowledge is as old an activity of the human mind as reflection is. One thinks here of the systems of Aristotele, Leibniz or Linnaeus (cf. Furner 2020; Kedrov 1975). Ordering systems are deeply embedded in philosophical systems and appear in all domains of human knowledge. We classify and order knowledge as we grow up in our individual ontogenesis, and we classify and order knowledge as mankind to enable orientation, navigation through growing masses of information. The ordering of knowledge is a precondition to allow for the abundance of new ideas by endless recombinations, alterations of marked cornerstones of human insights, next to flipping and breaking them and in this way pushing knowledge to a next level. In short, ordering knowledge is a prerequisite to linking knowledge.

The importance of organizing knowledge is based on its role in the functioning of the human brain. When we speak about organizing knowledge, we often refer to instruments, tools and principles that enable communication and the spread of knowledge beyond the individual. Systems (called here Knowledge Organization Systems or KOSs) are created to coordinate cognitive, communicative and social activities at the level of parts of society. Knowledge Organization is also the name of a scientific field, which grew from roots in philosophy, and emerged (arguably as a subfield in the information sciences) in the last century.

## 2.0 The oxymoron of abundant yet invisible KOSs

We are neither the first nor the last in struggling with ordering knowledge and with the question of how to best describe, reflect, teach and coordinate the practices of organizing knowledge. Each explosion of information almost naturally comes with visions and visioneers to make the best of large bursts of information. The emergence of the dedicated scientific domain Knowledge Organization (KO) is indispensably linked to the growth of knowledge by the last industrial revolution (end of the 19th, beginning of the 20th century), and the specific role of libraries in organizing it. In fact, classification theory (cf. Ranagathan 1973), as we know it today, had its origin in this time. New waves of technological evolution bring with them new challenges in ordering knowledge. For the information age, obviously the emergence of computers, and the emergence of networks of computers (the internet) are key.

The rise of automatization, based on computers as machines, goes together with formalization and abstractions, and KOSs of all kinds of flavour. There is no machine operation possible without strict formalization. Any data model (be it expressed in a database or a knowledge graph) relies on categorization and the definition of relations between them. Still, the more we depend on KOSs, the more they influence our lives, the less we seem to be aware of them as though the old dream of seamlessly supporting and guiding an envisioned user to information needs has become too perfectly realised. So, we experience an oxymoron: KOSs are everywhere and at the same time they seem to become more and more invisible.

To give an example from the world of scholarly communication: we have access to all libraries in the world almost, but in the design of the online public access catalog (OPAC), systematic catalog elements are often made invisible. For many purposes, their natural ordering function might be irrelevant. But, this is not always true. In particular, if one needs some context, a first orientation, an idea about which body of knowledge to consult, for those cases making existing classifications visible might be useful. But, there is more to the vanished ordering principles. Searching collections on-line is very different from visiting a physical place where a collection is held. In the physical world, the space alone conveys information. But, in front of a text-box, we often have no clue how large a collection is, how established the collection holding institution, or how this collection came about? There have been attempts to use the power of visualization to counter for that (Whitlaw 2015; Mutschke et al. 2017).

But, such invisibility of underlying KOSs does not only concern access to public or scholarly knowledge. A lot of our administrative, societal processes rely on data models and processing around them. Think about your nationality and the rights and obligations following from them. Here, categorization can deeply affect your existence. The problem augments with the mastery of artificial intelligence. We can process unprecedentedly large amounts of information, but we are concerned that we cannot even identify the algorithm principles behind them anymore, not to mention having an overview of consequences a certain design of data models and algorithms might have. To summarize, we are partly governed by hidden KOSs. This is an unsettling thought, when we remember that KOSs are always also a mirror of what is at stake in a certain society and culture at a certain time. (van den Heuvel and Zamborlini 2021, chapter 6 in this book). They are means of executing power by coordinating what to think, how to think and how to arrange access to the products of the mind (Bowker and Star 1999). They are not neutral. How can you find a position towards them, if you don't know them? To make KOSs visible and to revoke a discourse on the power of KOSs is the one aspiration of this book.

## 3.0 Orientation in the expanding KOS universe
Let's be clear: there is neither a way nor a reason to stop the information avalanche. Information is at the bottom of the knowledge-based economy (Leydesdorff 2006). As part of this avalanche, we as human beings will continue to organise knowledge as part of engaging in new practices of knowledge production. As KOSs are tailored towards specific practice, the more our societies differentiate and specialise internally, the more we will see different KOSs emerge naturally. So, we have to deal with this expanding KOS universe.

The rise of a great variety of KOSs can be compared to the sudden rise of innovations in certain ages, which in turn has been compared with the explosion of variants in biological evolution (Ziman 2003). Emerging new variants of KOSs can be described as mutations in such an evolutionary systems theoretical framework. But, from evolutionary theory we also know about a related thread, which has been called "mutation catastrophe" or has been referred to as "evolution window" (Rechenberg 1994). With too many variants existing, a comparison among them followed by a selection almost becomes impossible, all variants survive somehow and so evolution stops. Currently, it seems that we tend to build our own KOSs without even being aware of possibly useful already existing KOSs built by our neighbors. So, we seem to have given up on comparing and selecting also. Are we stuck in a "mutation catastrophe" of KOSs? That would be opposite to visions of Tim

Berners-Lee and others to truly connect and integrate knowledge (Berners-Le, Hendler and Lassila. 2001). So, how can we find a good balance between creating new KOSs as part of new practices and (re-)using, linking back to existing KOSs?

This question materializes not only on the level of KOSs. As the science system grows and new knowledge domains emerge at the boundaries of existing ones, there appears the need to foster, govern and organize interdisciplinarity. There is another scientific domain that formed its own epistemic framework, in particular after World War II, and this is known as Science and Technologies Studies (in short STS), a field at the crossroads of philosophy, history, sociology of science (here meant as the whole of academia) and innovation studies (Felt et al. 2017; Scharnhorst Börner and Besselaar 2012). A KOS in a knowledge domain (be it in academia or in any other societal sector) can be seen as a formalised expression of the epistemic framework, the conceptual reference, in which a domain operates. Thus, in a domain KOSs act as multipliers, coordinators of a certain worldview. Questions of how best to exchange knowledge between different domains cumulate in questions of how to bridge between different KOSs organizing those domains. The latter is part of the research in the field of KO, a field that also reflects on KOS design and that developed generic KOSs, which could govern this knowledge exchange (Smiraglia 2014b, Gnoli 2020). But, there are not many researchers combining methods from STS and KO (Smiraglia 2014a). For this book we selected chapters that combine presentation of knowledge ordering practices with a reflexive layer on those same practices.

**4.0 Linking knowledge by Linked Open Data**
The technological evolution of the last decades has not only fostered the emergence of ever more KOSs, it provided at the same time means to govern this explosion, namely SW technologies. As stated in Wikipedia (2020) "The Semantic Web is an extension of the World Wide Web through standards set by the World Wide Web Consortium (W3C)" with the ultimate goal of making information, data and knowledge machine readable. The visioneers, designers and engineers behind those SW-based technologies can be directly compared to pioneers in KO such as Paul Otlet and Henri La Fontaine who aimed at a universal bibliography containing all knowledge of the world and designed a system—a KOS, the Universal Decimal Classification—to navigate it (Rayward 1990; Wright 2014).

Equally, we see KOS at the heart of the Linked (Open) Data paradigm. According to Berners-Lee's (2006) 5-star system of data, linking documents on the web is only the first step toward deep linking of human knowledge (cf. Hyland et al. 2013). To enable access to the knowledge in documents is the essence behind "indexing" at the level of data. This takes the dream of Otlet and La Fontaine to a next level. Not only the works in which knowledge is documented (from music sheets to whole books, from images to technical drawings) are classified, but the "elementary units" in them such as concepts or phenomena should become (machine) referenceable. To be able to really weave those emerging network/graph representations of knowledge into one fabric of knowledge, it is necessary to also formalize the links between different KOSs (cf. Yoose and Perkins 2013). Among the ingredients that are part of this heroic effort we find: standards for expressing nodes and links in the knowledge graph and good practices (Hyland et al. 2014); an overview of existing KOSs in the LOD Cloud (cf. Linked Open Data Cloud; Vandenbussche et al. 2017); and requirements between the translation of KOSs (independently where they are published right now) into "semantic artefacts" (Le Franc et al. 2020) in a machine-readable form adhering to FAIR principles (FAIR stands for Findable, Accessible, Interoperable and Reusable [see Wilkinson et al. 2016]).

There is an abundance of literature about how to realise the LOD paradigm and which role KOSs play in this process (cf. Heath and Bizer 2011; Antoniou et al. 2012; Hyvönen 2012). Still, there seems to be a broken line of communication between specialists from different domains who reflect about the more generic aspects of KOSification. Those who reflect about classification theory and those who reflect about knowledge representation and reasoning are not always aware of each other's traditions, epistemic frames of reference and available solutions. As a consequence, there is quite some confusion about, for instance, what concepts are and how best to express them; and if natural language or classification languages and controlled vocabulary are better suited for expressing concepts (see Slavic, Siebes and Scharnhorst 2021; chapter 5 in this book). Some experts prefer to discuss these issues in the realm of library classification, others in the realm of computer science and engineering. In both camps there is much talk about the role of the user, however the role of humans in the engineering designs and of human use of machine readable KOSs is less clear. In general, the interplay between machines and humans around KOSs remains somewhat foggy. What can machines do and where are humans indispensable? If such questions are not properly sorted out among the information specialists and professionals, they create even more uncertainty among those applying the new knowledge ordering machinery in their daily research practice. To strengthen the link between different bodies of knowledge about KO and current KO practices in the realm of the SW is another aspiration of this book.

## 5.0 Bridging by reflecting

The main approach of the Digging into the Knowledge Graph (DiKG) project was reflecting by engaging in new practices. This was supported by bringing together experts with the various backgrounds referred to above. Among them were Wouter Beek, one of the designers of the *LOD Laundromat*, a "cleaned, indexed version of the Linked Open Data Cloud" (Beek et al. 2014). A further member of the famous knowledge representation (KR) group at the Vrije Universiteit Amsterdam[2], to which Wouter Beek belonged, Ronald Siebes worked during the project at the KNAW-DANS collaborating partner of the DiKG consortium. Richard Smiraglia, Rick Szostak, Daniel Martínez-Ávila, and Aida Slavic represented the KO experts. Andrea Scharnhorst contributed from the side of STS, DH and complexity theory. Together, we focussed on two areas:

•The KO discourse inside of the science of information and applied librarianship; and,
•The application of SW practices in parts of the digital humanities.

We started the call for this book with the statement (Smiraglia and Scharnhorst 2019):

The growth and population of the Semantic Web, especially the Linked Open Data (LOD) Cloud, has brought to the fore the challenges of ordering knowledge for data-mining on an unprecedented scale. The LOD Cloud is structured from billions of elements of knowledge and pointers to knowledge organization systems (KOSs) such as ontologies, taxonomies, typologies, thesauri, etc. The variant and heterogeneous knowledge areas that comprise the social sciences and humanities (SSH), including cultural heritage applications, are bringing multi-dimensional richness to the LOD Cloud. Each such application arrives with its own challenges regarding KOSs in the Cloud.

We also solicited contributions with a specific nature:

Working from the international "Digging Into the Knowledge Graph" LOD-KOS project (http://di4kg.org/)—we aim to bring together research papers from some of the world's leading experts in the application of multi-dimensional KOSs to the LOD cloud.

Next to the multidimensionality of KOSs we called for, the final contributions highlight the relevance of practices. Analysis of practices offers a way to identify barriers in knowledge exchange. The contributions o together represent a 'trading zone' in its own rights. In STS, Gallison's concept of a trading zone describes an intellectual, social-communicative place where different epistemic perspectives meet, exchange and through this exchange lay the ground for the emergence of new ideas, innovations and possibly new fields (Galison 1997). Indispensable for such a process is an openness towards talking about one's own implicit epistemic norms and values in a way in which others can relate to it. This goes together with awareness about boundary objects: concepts that change meaning when used in different knowledge domains, and which still can serve as a carrier for mutual understanding (Star and Griesemer 1989). The very definition of a KOS is a prime example for a boundary object (see Zeng and Mayr 2021; Chapter 3 in this book).

We designed this book as an intervention to current practices, advocating the need of a specific reflection layer and the acknowledgment of temporality in our endeavours. Too often, in particular in the early stage of the adoption of a new technology or method, explorations are presented as solutions ready to be re-applied. We wanted to counter this understandable but partly misleading attitude by unfolding that both the actions of "making a KOS" as an instrument for better research and "applying existing KOSs" represent research in its own right. It might look like an accompanying, supportive information management task, but the design of an information layer for a knowledge domain and the design of a domain specific research process are intrinsically interwoven. Still, at the same time, both processes require slightly different skill sets. We extended our intervention to KOS practices into education, by presenting a translation of existing standards and recommendations about LOD publishing for non-experts (Siebes et al. 2021; Chapter 12).

The book contains chapters that describe the state of the art in providing KOSs as semantic artefacts or semantification (Zeng and Mayr 2021, Chapter 3; Siebes et al. 2021, Chapter 12); how the state of the art is applied in new fields (van den Heuvel and Zamborlini 2021, Chapter 6; Smiraglia and Szostak 2021, Chapter 7; Patuelli 2021, Chapter 8; Smiraglia, Young and van Berchum 2021, Chapter 9); how the state of the art is pushed towards new technological solutions by being confronted with new applications (Oelen Stocker and Auer 2021, Chapter 10); how best practices need to be tailored towards specific solutions (Slavic, Siebes and Scharnhorst 2021, Chapter 5); and what challenges occur when merging new and old ways of expressing KOSs (van den Heuvel and Smiraglia 2021, Chapter 11; Szostak and Renwick 2021, Chapter 2).

To summarise, in this book the contributors address the problem of linking knowledge in two different ways:

•To address fundamental issues of KO: such as presentation of concepts, roles of different KOSs (thesauri, analytico-synthetic classifications) and their representation as Linked Open Data
•To make the role of KOSs and the practices behind the design of KOSs visible in areas of scholarly communication and certain fields of humanities research.

The reader might ask, why did we choose the rather traditional form of an edited collection to document the results of this research? The main reason is the relative stability of written documentation. Of course, the DiKG project (as well as the other projects that contributed to this book) also delivered other kinds of output: LOD, architectural designs, experimental services, and a large amount of Resource Description Framework (RDF)-modelled content. But, SW technologies, as mentioned earlier, represent also a fast moving research front with ever new approaches and corresponding tooling. As often discussed by the community itself, sustainability of published resources and solutions is an issue (cf. Benjamins et al. 2002). This can best be illustrated with the case of the LOD Laundromat. At the beginning of the DiKG project, in 2017, the LOD Laundromat (Beek et al. 2014) was still in operation. Back than one could find a whole suite of tools provided to search the crawled LOD cloud (see https://web.archive.org/web/20190103031340/http://lodlaundromat.org/ for the landing page) Currently, the website http://lod-a-lot.lod.labs.vu.nl serves as a new experimental space for researchers from the VU Amsterdam group, but the original tools are no longer available. It is important to note that the content of the LOD Laundromat was archived (albeit as large dump of static linked data) with a long-term archive (DANS-EASY) (Beek et al. 2017). Still, the cleaned LOD version is no longer available as a service. The LOD Laundromat is not the only web-based resource that has experienced such life cycle changes. Also, the CMME web-resource that we used in the DiKG project changed in functionality during the project's life-time. The chapter by Slavic, Siebes and Scharnhorst (2021; Chapter 5) shows in detail the care a service provider has to take when introducing new forms of a service. In many cases, researchers or research projects are the owners of linked data (LD) solutions, and often they are described as workshop, lab or experimental material and not designed to operate as sustained production services. In order to preserve the efforts that go into creating KOSs and KOSs as LD a close collaboration with institutions that guarantee stability is needed. As illustrated in the chapter by Smiraglia, Young and van Berchum (2021; Chapter 9), publication of LOD can often also take the form of submitting content to services maintained by others. In this line of reasoning, even archives have a role, as LD in the form of RDF is no more than a very detailed "index" or description. Even if the machinery to execute operations with this index is no longer functional, it still makes sense to document and preserve the efforts behind the RDF modelling and indexing of content with it. But, of course, documenting is wise also in competition with actual executing research, and no detailed research data management strategy will ever be able to solve this dilemma.

This brings us to a last disclaimer. While we argue in favour of linking new and old cultures of documentation and KO, we put special emphasis on making time and space for reflection in explorative practices. We see reflection in processes of cross-domain communication as an indispensable means to achieve (better) results. However, we are very well aware that there always remains a tension, the tension between making and analysing; between pushing technology forward and applying existing technology; between being precise, well rooted in your own domain and reaching out to other domains. We started this introduction with an emphasis on the key role KO in general and KOSs as specific instruments. So, the bar is high. But, even here, one needs to find a balance between the efforts requested and the benefits expected. With KOSs the situation is no different from more general discourses around data. For instance, in designing new IT services for data search and sharing one needs to balance the costs with the expected benefits (cf., Gregory 2021). While, there is endless potential in making all KOSs semantic by curating and observing them, we might not be able to achieve this. No, this is not a call to give up; but a call to be aware of it and to honestly present achievements together with limitations; to be explicit about what should, could, and what most probably will be realised.

## 6.0 Content of the book
### 6.1 Overview
The book is organised in five sections: ***Background, Foundations, Applications, New Endeavours, and Education***. In the overview, we list all chapters in those sections, and proceed further to summarise what those chapters bring to the book in the light of the goals of this book as discussed above.

#### *Background*
Andrea Scharnhorst and Richard P. Smiraglia (Chapter 1)
> "The Need for Knowledge Organization: Introduction to the book *Linking Knowledge: Linked Open Data for Knowledge Organization*

Rick Szostak, Richard Smiraglia, Andrea Scharnhorst, Ronald Siebes, Aida Slavic, Daniel Martínez-Ávila and Tobias Renwick (Chapter 2)
> "Classifications as Linked Open Data: Challenges and Opportunities"

#### *Knowledge Organization and Linked Data - Foundations*
Philipp Mayr and Marcia Zeng (Chapter 3)
> "Knowledge Organization Systems in the Semantic Web: A Multidimensional Review"

Tobias Renwick and Rick Szostak (Chapter 4)
> "A Thesaural Interface for the Basic Concepts Classification"

Aida Slavic, Ronald Siebes and Andrea Scharnhorst (Chapter 5)
> "Publishing a Knowledge Organization System as Linked Data: the case of the Universal Decimal Classification"

#### *Application of Linked Data in the Digital Humanities*
Charles van den Heuvel and Veruska Zamborlini (Chapter 6)
> "Modeling and Visualizing Storylines of Historical Interactions: Kubler's *Shape of Time* and Rembrandt's *Night Watch*"

Richard P. Smiraglia and Rick Szostak (Chapter 7)
> "Identifying and Classifying the Phenomena of Music"

M. Cristina Patuelli (Chapter 8)
> "Graphing out Communities and Cultures in the Archives: Methods and Tools"

Richard P. Smiraglia, J. Bradford Young and Marnix van Berchum (Chapter 9)
> "Digging into the Mensural Music Knowledge Graph: Renaissance Polyphony meets Linked Open Data"

#### *Knowledge Organization and Linked Data - New Endeavours*
Allard Oelen, Mohamad Yaser Jaradeh, Sören Auer and Markus Stocker (Chapter 10)
> "Organizing Scholarly Knowledge leveraging Crowdsourcing, Expert Curation and Automated Techniques"

Charles van den Heuvel and Richard P. Smiraglia (Chapter 11)
> "Knowledge Spaces: Visualizing and Interacting with Dimensionality"

#### *Knowledge Organization and Linked Data - Education*
Ronald Siebes, Gerard Coen, Kathleen Gregory and Andrea Scharnhorst (Chapter 12)
> "Publishing Linked Open Data: A Recipe"

## 6.2 Background

In the background section the reader finds next to this introduction, a reprint of the paper "Classifications as Linked Open Data: Challenges and Opportunities" (Szostak et. al 2020). This paper summarises the achievements of the DIKG project. It discusses in particular the challenges that emerge when classifications designated for the bibliographic domain, be they of newer or older provenance, are prepared to be interwoven into the LOD Cloud. The cases of the Basic Concept Classification (BCC) and the Universal Decimal Classification (UDC) will be discussed in further detail in chapters in the next section.

## 6.3 Foundations

The foundations section starts with the chapter "Knowledge Organization systems in the Semantic Web: A Multidimensional Review." Zeng and Mayr unfold how complex and to a large extent not yet fixed the terminology is when it comes to questions of "what is a KOS?" "how a KOS as a model of knowledge can be made machine readable" and "how KOSs can be used in machine readable statements." What makes this contribution so special is that it sheds light on the different actors involved in making, providing and using KOSs in the context of the SW. The authors do this by designing personas or proto-personas, an approach from experience design. One large group of those personas is the providers of LD services of KOS. Providers can operate country-wide or deliver just one individual vocabulary. Services providing KOSs as semantic artefacts (Le Franc et al. 2020) also include middleware for end-users or registries. The KOS service providers are just one part of the wider landscape of consumers and producers of KOSs. Similar to what has been found for users of data (Borgman et al. 2019), and in many other studies on knowledge production practices (cf. Wouters et al. 2013) roles are usually mixed in practice. So, one and the same organization, group or person can operate as different proto-personas depending on their actual activities. Next to the service providers Zeng and Mayr introduce the dataset producer using LOD principles, the vocabulary producer, research groups as end-user, website and tool developer. All of them can operate on various geographical levels and in or across different knowledge domains. The authors derive those archetypical personas from a rich empirical analysis of the field as it stands now. Two aspects here are striking: first how experimental the stage of semantic KOSs still is, how fluid, how much in development (this is, by the way, a thread through all the chapters in this book) and second the gap between makers of KOSs and makers of KOSs in LOD form.

The second chapter "A Thesaural Interface for the Basic Concepts Classification" (reprint of Renwick and Szostak 2020) discusses how to design an interface that can guide a human executing a classifying task (indexer or classifier) through the BCC controlled vocabulary. The paper departs from fundamental issues of language-based classifications and tries to bridge between keyword and subject search activities by the design of an interface. The final aspirations of this use case of the DIKG project are higher, namely to support user queries formulated in sentences.

The last foundations chapter, "Publishing a Knowledge Organization System as Linked Data: the case of the Universal Decimal Classification," documents the efforts of a KOS service provider to make one of the standard bibliographic KOSs available as LD. The Universal Decimal Classification (UDC) is an analytico-synthetic and faceted classification whose origins go back to the end of the 19th century. Paul Otlet and Henri La Fontaine started in 1896 an international project intended to cover all information sources published in human history, in any form or language, anywhere in the world (Wright 2014). The UDC design as a synthetic indexing language and its use in practice over a long period of time has not only influenced further design improvements of the UDC (Smiraglia et al. 2013; Slavic and Davies 2017), but contributed to the large amount and variation of UDC codes and their combinations in bibliographic metadata (Scharnhorst et al. 2016). How, in general, KOS expressions are created by local practices when a specific KOS is applied is a complex process in itself (Tennis 2012). It is also beyond the control of KOS editors and publishers. But, the KOS service providers have to address the large user base of their KOSs, and this problem augments, if both KOS service and KOS use become part of the LOD cloud. For bibliographical metadata we can observe that Machine Readable Cataloging Records[3] increasingly become available as LOD. In the case of the UDC, to further be able to use its analytical power, it is necessary to build the UDC as a semantic artefact (Le Franc et al. 2020) in a way that preserves both the structure of the UDC and its provenance over time. Slavic, Siebes and Scharnhorst present a detailed discussion of those issues and unravel how those influence the final architectural design for a new LD/LOD publishing service of the UDC.

## 6.4 Applications

As indicated above, this book discusses not only generic issues of KO when it comes to LOD, but it also zooms into practices currently applied in the DH domain. In the applications section two areas are covered: KO and knowledge graph designs in the prestigious Golden Agents[4] project, and classification issues around works and practices in music and musicology.

Van den Heuvel and Zamborlini contirbute "Modeling and Visualizing Storylines of Historical Interactions: Kubler's *Shape of Time* and Rembrandt's *Night Watch*." They describe how discourses and controversies in art history come to new life when building a knowledge graph that enables weaving different historical sources into one information fabric. One aspiration of the Golden Agents project is to be able to understand what we would call today "creative industries" during the Dutch Golden Age (ca. 1581-1672) in their entirety, covering different sectors and their different products, and describing the role of different actors (producers and consumers) (cf. Idrissou et al. 2019). Biographies of makers, networks of their interactions and traces of objects in space and time will all come together. Naturally, harmonization of information about agents and objects and dealing with a large variety of KOS standards used in the different branches of all knowledge domains involved are at the heart of this project. But, in this chapter, the authors focus on how temporality of events and processes should be captured in a way that allows for different stories about the past to appear. Naively, to pinpoint everything to an external arrow of time seems to be the obvious solution. However, the real challenge lies in the selection and later standardised description of what to connect to which point in time in which way. Processes come with their own temporal signature, objects can be found in different manifestations, stories about both deliver additional information but again come with their own temporal provenance. The chapter details how a model emerges that allows describing this complexity in a way that machine-based information processing as well as linking to other sources becomes possible. Here the emphasis is on standardization and formalization. A better retrieval of information is a very tangible outcome of the project. Yet, and this is the real focus of this chapter, the model should also be flexible enough to still support the hermeneutic, interpretative, explorative research practice that produces new, fresh insights beyond agreed standards. In discussing the decisions behind the eventual chosen model a trading zone for concepts, epistemics frames and language to describe them becomes visible.

The next paper, "Identifying and Classifying the Phenomena of Music" (reprinted from Smiraglia and Szostak 2020) continues the discourse around how to best (for certain purposes) represent artifacts from the past. Smiraglia and Szostak focus on the phenomena of music, and call for an extension of the usually documented features in music retrieval. This discussion is actually based on mimicking or envisioning future research behaviour of musicologists. As always in the history of documentation, documentation of resources for research goes hand in hand with existing (but not answerable) and newly envisioned research questions. New designs for KOSs should be best on an analysis of current research practices in a certain knowledge domain. But indexers base their indexing on the content of the work as well as on their imagination of a use and a user. The KO providers (see Zheng and Mayr 2021; Chapter 3) the designer of KOSs that only partly overlap with those later using the KOS are both faithful to observations and visionary. Smiraglia and Szostak use a generic classification, the BCC, to identify facets that might become candidates for further standardised documentation.

The next chapter, "Graphing out Communities and Cultures in the Archives: Methods and Tools," puts such considerations in action. Patuelli describes the project *Linked Jazz*[5], which explores the power of LOD technologies when applied to the history of jazz. Similar to the Van den Heuvel and Zamborlini (2021; Chapter 6), the quest is again for a web that connects entities, people, objects, facts and concepts in new and unprecedented ways, across disparate domains and beyond repository boundaries. More specifically, the chapter (and the project) uses an oral history approach departing from existing interviews with jazz musicians in certain repositories. Using the power of the Wikidata platform, a network of information emerges that allows different perspectives, such as the social ego network of an artist, or the location history of the emergence of certain genres and styles.

Pushing boundaries of existing KOSs or recombine existing KOSs into something new have been topics of KO through history. Equally, sustaining insights inscribed in KOSs has been achieved by institutions issuing authoritative resources, and by guarding workflows around those authoritative resources. "Digging into the Mensural Music Knowledge Graph: Renaissance Polyphony meets Linked Open Data," is a description of just this. An important contribution concerns the enrichment of existing information about composers, works and sources in the Virtual International Authority (http://viaf.org/) by incorporating information from a very specific research-driven curatorial project about mensural music. But, Smiraglia, Young and van Berchum (2021; Chapter 9) also proudly report that for the first time (173) "a corporate cultural heritage entity that was not a cataloging library [has] been allowed to participate in LC/NACO [Library of Congress]/… to enter authority records directly." It is by such pathways for integrating digital humanities results with authoritative, stable (KOS) service providers that sustainability becomes most effectively achieved. The mensural music project also serves to demonstrate that producing LD does not always mean to lift the whole of a resource to the LOD cloud; but rather that one can select different levels when it comes to the process of LOD publishing (Siebes et al. 2021; Chapter 12). Again, it very much depends on what your role in the KOS universe might be (Zeng and Mayr 2021; Chapter 3).

## 6.5 New endeavours

The new endeavours section sheds light on two specific dilemmata in current linking-knowledge practices. The first concerns the co-evolution of a technology and social practices of its adoption. Automatic indexing is ever evolving towards finer granularity, from indexing documents (or websites) to indexing bits of content in documents in even more precise ways. SW technology is just the fuel to enabling linking in unprecedented larger (all encompassing) and deeper (more detailed) levels. Allard Oelen, Mohamad Yaser Jaradeh, Markus Stocker and Sören Auer have contributed chapter 10 "Organizing Scholarly Knowledge leveraging Crowdsourcing: Expert Curation and Automated Techniques." They have addressed the question of how to enable indexing on the level of methods, so that in turn, a user can more quickly gain an overview about methodological achievements. In the process of KOS creation and application for automatic indexing experts are still indispensable. Machines might be able to suggest a KOS structure, but validation requires human intervention. It is up to the users based on their best practices to suggest, select, test, apply and re-design the KOS in question repeatedly. In this sense, this chapter once more re-emphasizes the position of the user in machine-based/automatic classification.

The dream of a knowledge graph that enables a new organization of human knowledge—the old Otlet/La Fontaine dream so to say—also encapsulates another dilemma. To be able to scale up, to connect across knowledge domain boundaries, one needs to identify elements that are generic and can act as bridges between those domains. Inevitable, that means that concreteness, semantic embeddedness in one specific context will need to be washed out. In other words, one has to find a formalization that enables the trading zone discourse, which we have so often pointed to in this introduction. This dilemma—generic versus specific—is an eternal problem, and answers to it eventually relate back to fundamental philosophical stances: is it ever possible to find common ground or are we doomed to be stay confined inside our own individual, local world views? Is there an objective reality, and if so how do we gauge and evaluate different representations of it over and against each other? The last chapter is this section (Chapter 11) is by van den Heuvel and Smiraglia, who address exactly these questions seeking a conceptual framework for thinking about possible solutions. In "Knowledge Spaces: Visualizing and Interacting with Dimensionality," they discuss how best to enable access to different perspectives, or representations. While socially constructed, such perspectives are not arbitrary. The authors go one step further and discuss how to turn the problem (the existence of different perspectives) into a solution (enabling a better understanding). "Ordering the Ordering Systems" and "Using Visualizations" are part of their answer. In essence, they seek (201) "a more instrumental use of multidimensional knowledge spaces to organise and to interact with concepts."

## 6.6 Education

The last section of the book brings focus to the aforementioned dilemma resulting from the co-evolution of technology and its use in a different way. The virtue of SW technology—its innovative character and ever-new emerging possibilities—can turn into a fault when it comes to its adoption, and in particular when the adoption is not properly managed. One has to acknowledge that when operating at the boundary of different fields, investment needs to be made in the translation process concerning concepts and approaches as well as eventual experimentation and implementation. Concerning LD one can observe all kinds of myths circulating. There is sometimes a naive belief that once data are transferred to a LOD format they automatically become interwoven into larger knowledge graphs, and sometimes the same belief can transfer into fear of losing control over one's own data, argument or research. Both beliefs arise from incomplete information, and the inhibit further wider adoption. The professional organization W3C[6] of the SW field creates extensive documentation to foster standardization processes. But, their main addressants are the experts and professionals in the field, not primarily adopters from other fields. So, while there are ample recommendations on how to publish LOD, there is still a need for educational material. This observation, made in many projects, motivated Siebes, Coen, Gregory and Scharnhorst to engage in a so-called "sprint" organised by the Mozilla Library Carpet movement and to write a guide for LOD publishing for everyone. Based on existing W3C recommendations, steps (called "things" in the Library Carpet format) were identified that need to be pursued in bringing an information resource to the LOD cloud. This guide also informed the architectural design of a new L(O)D service of the UDC (Slavic Siebes and Scharnhorst 2021; Chapter 5). However, in applying the guide to a use case of our own we once more experienced that formal workflows (as such a guide represents) in research practice are really thinking tools rather than automatic fabrication tools. In each application context, those steps to follow will be different. Having said this, one important message of this chapter remains: namely many steps in the process of publishing LOD require thinking prior to programming, and could be executed with pen and paper. In other words, each LOD project needs a blueprint as well as a machine. To produce the first, "translation work" is needed, to produce the second, a specialist from the SW community is needed.

## 7.0 Linking knowledge: The synergy of knowledge interaction

The title of this book was not arbitrary. Not a bit. The explorers who have collaborated in this book are a team devoted to moving beyond the simple concatenation of RDF triples into a realm where the linking of data represents a true linking of knowledge, which itself becomes a linking for knowledge interaction. We all set out on a journey to make sense of the chaos in the World Wide Web, no less than did our predecessors over the past three centuries try to make sense of the chaos unleashed by the printing press. We have laid out a path much more useful than the cookie crumbs of Hansel and Gretel. We have followed a path set out for us over centuries of work on bringing to fruition the most possibly useful organization of knowledge. We humbly present this book as evidence of our journey.

The useful linking of knowledge across spectra is an eternal human dream. From ancient stargazers to scientists of the LOD Cloud, like all those who have contributed to this book, the goal is the realization of two mid-20[th] century scholarly dreams: facilitating what Patrick Wilson ([1968] 1978) called exploitative power, or the power to synthesize knowledge with laser-like precision; and to do so by bringing together what Don Swanson (1986) called undiscovered public knowledge, in other words facts related in as yet undiscovered ways. These two goals meet in what van den Heuvel and Smiraglia (2021; Chapter 11) call "knowledge interaction." The rise of the idea of the SW represents the realization of these two dreams in much the same way as the automation of bibliographic control in the last quarter of the 20[th] century led to one realization of the Otletian dream of universal control of research, first through bibliographic utilities like OCLC, Inc., and then subsequently through the rise of the World Wide Web. That is to say, while the SW promises us the power of both exploitative ability and unfettered synthesis, still it also represents a divergence from the imaginable technologies of the past and therefore is a pathway to new and as yet unimagined exploration. The authors contributing to this book—our explorers— have reported their observations and wisdom concerning the forging of this new dream, with the especially delicious twist of a focus on the social sciences and humanities (SSH).

Along the way our explorers have discovered and here reported on the parameters of what some of them call a new ecosystem. Pattuelli (2021; Chapter 8) refers explicitly to the "linked data (LD) ecosystem." The ecosystem is bounded by RDF triples, which are themselves the outgrowth of a universe of data. Similarly, this ecosystem is populated by realized knowledge infrastructure in the form of knowledge organization systems (KOSs) that Zeng and Mayr (2021; Chapter 3) refer to variously as communities, researchers, producers and users as well as colonies. Van den Heuvel and Smiraglia (2021; Chapter 11) relate this populated ecosystem to Otlet's visualizations of the dichotomous "Self (le Moi)" and "Societies (Societés)" as coexisting realities of perception of the organized knowledge universe. What kind of thing, then, is this new SW reality?

Smiraglia (2014a) wrote about the potential synergies of information institutions as social realities. Information institutions are defined as those (1) "that preserve, conserve and disseminate information objects and their informative content." The commonality among information institutions lies in (2ff.) their shared mission to disseminate knowledge by means of some sort of query-response system, and that by virtue of these they manifest a form of *gravitas*. Cultural synergy "is the combination of perception- and behavior-shaping knowledge, within, between, and among groups that contributes to the now realized virtual reality of a common information-sharing interface." It seems obvious that a knowledge-sharing environment as rich and lively as the semantic linking ecosystem, populated and colonized by communities of researchers, producers and users both constitutes and is comprised of information institutions. The LD ecosystem(s), the LOD Cloud, the SW and their constituent LOD KOSs and LOD knowledge graphs (KGs) all qualify as information institutions. The synergies among them are the real thesis of this book.

There is a critical element that we can use to help us comprehend the evolving cultural synergies shaping the LD ecosystems and that is the notion of social epistemology (6): "information institutions arise culturally from social forces of the cultures they inhabit, and … their purpose is to disseminate that culture." Certainly, the interwoven layers of data ecosystems, but in particular LOD KGs, populated by the "societies" of communities of researchers, users and producers demonstrate in the action of colonization around "approved" (cf. Zeng and Mayr 2021; Chapter 3) LOD KOSs are entirely creatures of the cultures from which they have sprung and are determined disseminators of their culturally requisite knowledge stores. One synergy is immediately apparent and that is the intermingling of cultural realities of the LD producing community of computer and information scientists, on the one hand, and the rich SSH communities of researchers and users, on the other. Examples in this book are LOD KOSs such as the Basic Concepts Classification and the Universal Decimal Classification, the Golden Agents, Mensural Music, Linked Jazz and Open Research knowledge graphs. In each case the KG is the intermingled product of interdisciplinary interaction between the SW and SSH communities. The synergy is the dualistic social epistemology—these KGs are disseminators not only of their research content but of their constantly evolving SW ecosystems as well.

Specific synergies exist also in the infrastructural elements of the LD ecosystem. Vectors in knowledge space, described by van den Heuvel and Smiraglia (2021; Chapter 11), are essentially syndetic connectors that cross pathways intersecting not only specifically linked data but also the data ecosystems surrounding each such linkage. The vectors are synergistic vehicles navigating undiscovered related conceptual space creating knowledge interaction. The knowledge space in which the vectors operate is the synergistic multidimensional knowledge space of a universe of KGs, themselves connected epistemologically by the methods underlying their construction and by the very fuzzy nature—noted by Renwick and Szostak (2021; Chapter 4)) as well as by van den Heuvel and Zamborlini (2021; Chapter 6)—of SSH domains where research relies on inexact linkage to generate useful matches.

Another synergy is the power of the LD ecosystem to merge the historical record, including evidence of the products of creative action. These elements meet, indeed suffuse one another, in the LD world of the bibliographic authorities for designated creators (authors, composers, etc.) where the library driven Virtual International Authority File (VIAF) bumps against DBPedia, but lesser known creators ranging from jazz musicians to contributing librarians to artisans of the Dutch Golden Age are identified through crowdsourced Named Entity Recognition (NER) modules. Pattuelli (2021: Chapter 8) identifies the importance of this synergy by reminding us that (162) "the full potential of LD is reached when heterogeneous data from different sources are interlinked providing unified access to data and the possibility to seamlessly query multiple graphs."

Classification is perhaps the most powerful tool ever devised by science. Its emergence in the LD ecosystem as the queen of the LOD KOS is testimony to its virtue for both gathering and disambiguation. Divergent philosophies underlie potential universal (i.e., general) classifications and therefore their potential synergistic effect when used in combination in the LD ecosystem. The discipline-based UDC has the power of over a century of application in the linking of the documentary evidence of recorded knowledge (cf. Slavic, Siebes and Scharnhorst 2021; Chapter 5). A late 20th century competitor was the Information Coding Classification of Dahlberg (cf. van den Heuvel and Smiraglia 2021; Chapter 11), which is liberated from the constraints imposed on the UDC by replacing disciplines with ontical structures. The phenomenon-based Basic Concepts Classification (cf. Renwick and Szostak 2021, Chapter 4; Smiraglia and Szostak 2021, Chapter 7) is designed to promote interdisciplinarity by structuring phenomena in causal relation sequences. There is an emerging synergy produced by the use of any and all of these classifications (see Szostak, Smiraglia, Scharnhorst, Slavic, Martínez-Ávila and Renwick 2021; Chapter 2) not only as LD themselves but in conjunction with each other as descriptors linked to points representing concepts in the LD Cloud. It is as though each classification represents a distinct dimension in the knowledge universe. The points in each dimensional knowledge space representing classified concepts or phenomena become additional vectors crossing the many dimensions to create synergistic knowledge interaction.

Our intrepid explorers (the authors who contributed to this book) did not embark on this frontier unprepared. We can partially observe the manifold provisions for this journey by analyzing the discourse they share. Discourse analysis is an evolving methodology of domain analysis in KO, seeking identification of the conversation, or "discourse," to reveal underlying points of view shared by authors in a domain. According to Smiraglia (2015, 15): discourse analysis is one means of revealing the interacting symbolic contexts in the discourse that are affecting perception … [by] selecting key elements of discourse in a domain." Whereas other methods of domain analysis reveal the ontology at work in a domain, discourse analysis helps narrate the collective theoretical framework. Techniques for discourse analysis vary from ethnographic narrative analysis to informetric analyses. Smiraglia (2018) demonstrated the use of bibliometric analysis to reveal the contours of domain discourse. For the purpose of discovering the discourse present among the authors whose work appears in this book we have compiled and analyzed the reference lists from all twelve chapters.

For example, there are 351 references to works cited in the twelve chapters, of which 25 references are cited three or more times making up one third of all references. Not surprisingly, the most-cited authors are among the contributors: Smiraglia (25), Szostak (12), van den Heuvel (11), Slavic (10) and Pattuelli (6). Although there is some self-citation, which is common on a research front where the authors are reporting sequential new research, there also is a fair bit of cross citation. That is, these contributors know, rely upon, and perhaps most importantly acknowledge their reliance upn each other's work. Works cited three or more times give a clue to the community discourse. These are (Table 1):

| Authors | Title |
| --- | --- |
| | The LOD Cloud |
| Berners-Lee (2006) | "Linked Data" |
| Idrissou, Zamborlini, van Harmelen and Latronico (2019) | "Contextual Entity Disambiguation in Domains with Weak Identity Criteria: Disambiguating Golden Age Amsterdamers" |
| Rayward (1990) | *International Organization and Dessemination of Knowledge: Selected Essays of Paul Otlet* |
| Smiraglia and van den Heuvel (2013) | "Classifications and Concepts: Towards an Elementary Theory of Knowledge Interaction." |
| Szostak, Scharnhorst, Beek and Smiraglia (2018) | "Connecting KOSs and the LOD Cloud" |

Table 1. Works cited three times or more in this volume.

A very interesting backdrop to the shared discourse emerges. First, we have the actual living LOD Cloud, which is clearly in every mindset. Alongside that visualization of the SW we have two historical outposts—essays by Paul Otlet, the 19th century visionary who postulated something like a semantic universe that might be technologically feasible, and musings on the technicality of LD by the SWs own 21st century visionary Berners Lee. The three remaining works include an extensive theoretical essay on visualizing knowledge interaction (Smiraglia and van den Heuvel 2013), a paper on the essence of bringing data from the SSH into the LOD Cloud—the inexactitudinous nature of SSH data, which often requires the use of inexact matching for interpretation (Idrissou et al. 2019), and the opening salvo from the Digging Into the Knowledge Graph team concerning the necessity and processes for connecting the LOD Cloud to traditional KOSs (Szostak et al. 2018).

Author co-citation analysis is a technique by which all pairs of authors cited together in a domain are mapped. In general, co-citation indicates perceived association (e.g., semantic, thematic, epistemological, etc.) on the part of the citing author between the members of a pair. That is, a co-citation map shows how citing authors perceive associations among cited works. In domain analysis the technique is useful for visualizing theoretical poles in a specific domain, or we might also say nodes of discourse, represented by the perceived associations. Visualization takes place by using multi-dimensional scaling (MDS) to generate a network map of co-cited authors. Each referenced author forms a node in the network, and edges between authors occur if they are co-cited. The weight of the edge indicates the strength of the perceived association; thicker edges represent more frequent co-citation, thus revealing more influential theoretical poles. As part of our discourse analysis we plotted co-citation across the twelve chapters of this book; Figures 1 and 2 present two views of the author co-citation network in this volume.

Figure 1 shows a Gephi plot of author co-citation among those authors most cited in this volume. This gives a clue to the shared discourse, or conversation, among the citing authors, who are (of course) the authors contributing to this volume. Here we ask the question, what theoretical poles have influenced the work underlying the collective contributions to the idea of linking knowledge.

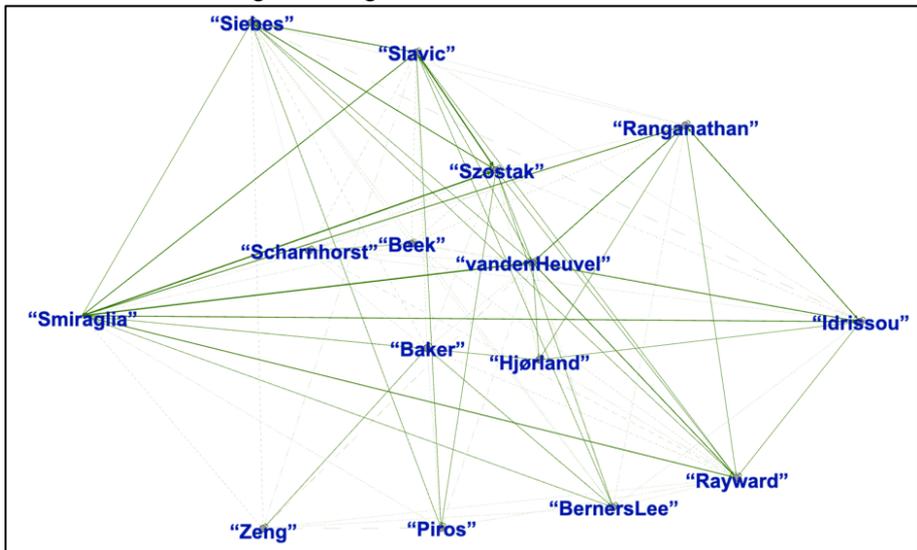

Figure 1. Author co-citation among those most cited.

At the core of this network we find a cluster including key contributors to this volume, but the core is informed by work by Ranganathan and Rayward (Otlet), which is evidence of the historical grounding of the discourse. Also interesting is the prominence of applications from Beek, Piros, Siebes and Idrissou representing the keen importance of the specific technologies necessitated for connecting KOSs and the SSH to the LD ecosystem. The strongest connection shown by the heaviest edges is the network among Smiraglia Szostak Slavic and van den Heuvel.

A slightly different view of the discourse can be generated by restricting the analysis to only those authors who are contributors to the volume. In other words, we now ask, how do these authors view each other's theoretical contributions to the notion of linking knowledge? Figure 2 shows a Gephi plot of author co-citation among authors contributing to this volume.

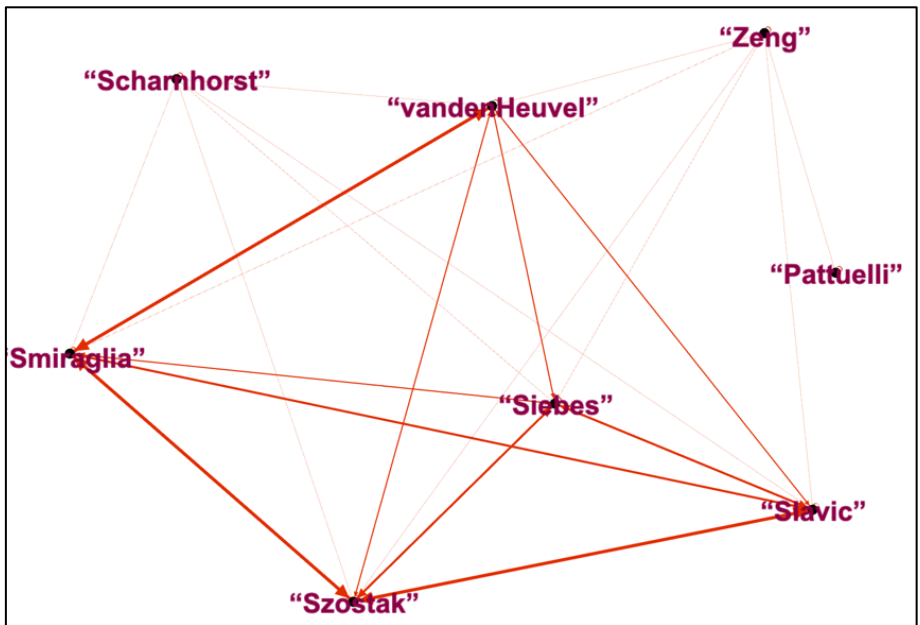
Figure 2. Author co-citation among contributing authors.

Obviously, authors from Figure 1 not represented in this plot were not co-cited in the volume, leaving a very explicit imprint of the loose but apparent discourse at work. Here we see the degree to which the core authors—our most intrepid explorers—rely on each other's work. The theoretical core overlaps that from Figure 1—the principal core related to the linking of knowledge in SSH via KOSs is anchored in reliance on specific LD technologies and buttressed by a strong connection to ideas about visualizing knowledge interactions.

Co-word analysis of the same dataset (the titles of the works cited in the volume) can be used both as a form of methodological triangulation and as a means of informing the interpretation of the discourse visualization. The Provalis ProSuite's WordStat module was used to help to visualize the core concepts represented in the research cited by our contributing authors. Figure 3 is an MDS plot of the most frequently occurring keywords and Figure 4 is a plot of the most frequently occurring two to five-word phrases.

Figure 3 shows the boundaries of the discourse at play: the core cluster is a combination of "classification" "information" and "knowledge" orbited by the SW, LD and historical memory. Figure 4 gives more breadth to the discourse by showing the core of SW LD and cultural heritage orbited by iterations of knowledge graphs and the fascinating cluster including a "universe of knowledge" and "information retrieval." We also see the prominence of the phrase "contextual entity disambiguation in domains." Thus, there is synergy in the discourse across historical and immediate imperatives driven by interdisciplinary approaches to knowledge interaction. That is the grace of this book.

Let us then recount the ways in which our non-arbitrary project has produced non-trivial synergies:

•Intermingling of cultural realities of the LD producing community of computer and information scientists and the rich SSH communities of researchers and users;

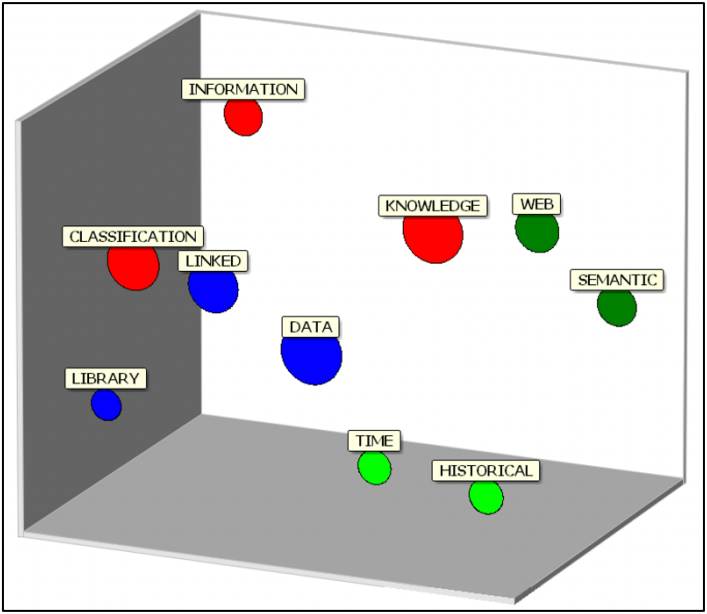

Figure 3. Most frequently occurring keywords (stress = .017138 $R^2$ = .9602).

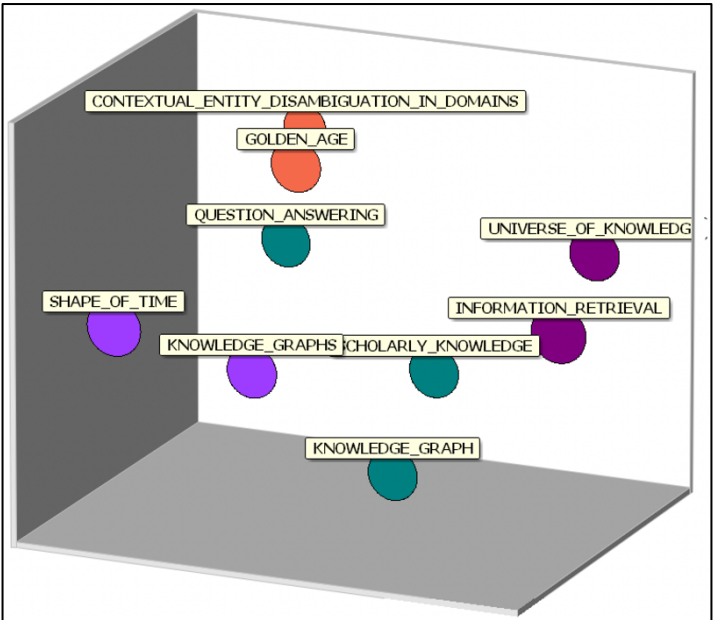

Figure 4. Most frequently occurring phrases (stress = .18684 $R^2$ = .9670).

•Vectors in knowledge space are synergistic vehicles navigating undiscovered related conceptual space creating knowledge interaction;

•The power of the LD ecosystem to merge the historical record, including evidence of the products of creative action;

•Divergent philosophies underlie universal (i.e., general) classifications and therefore offer a potential synergistic effect when used in combination in the LD ecosystem;

•There is an emerging synergy produced by the use of any and all of these classifications not only as LD themselves but in conjunction with each other as descriptors linked to points representing concepts in the LOD Cloud.; and,

•There is synergy in the discourse across historical and immediate imperatives driven by interdisciplinary approaches to knowledge interaction.

Ultimately the LD ecosystem explored and documented so eloquently by the contributors to this volume represents a potentially unbridled source of knowledge generation, acquisition, production and dissemination. The underlying discourse shows the extent to which our explorers are firmly grounded by historical vision yet equally firmly dedicated to the promise of linking knowledge for interaction. This new SW reality is an exciting frontier of fascination, expansion and growth. The already maturing ecosystems of the SW are interlocking information institutions clearly devoted to the expansion of human experience through the growth of knowledge interaction.

## Notes

1. Digging Into the Knowledge Graph (DIKG). https://digingintodata.org/awards/2016/project/digging-knowledge-graph
2. https://www.cs.vu.nl/~frankh/#
3. https://www.loc.gov/marc/umb/um01to06.html
4. https://www.goldenagents.org
5. https://linkedjazz.org
6. https://www.w3.org